\documentclass[12pt]{spieman}  % 12pt font required by SPIE;
\usepackage{amsmath,amsfonts,amssymb}
\usepackage{graphicx}
\usepackage{setspace}
\usepackage{tocloft}
\usepackage{multirow}
\usepackage[square, sort, comma, numbers]{natbib}
\usepackage{lineno}
\usepackage{xcolor}
\usepackage{float}
\title{Forecasting Wavefront Corrections in an Adaptive Optics System}
\author[a,b,*]{Rehan Hafeez}
\author[a,c]{Finn Archinuk}
\author[a,c]{Sébastien Fabbro}
\author[a,c]{Hossen Teimoorinia}
\author[a,c]{Jean-Pierre Véran}

\affil[a]{National Research Council Canada, 5071 West Saanich Road, Victoria, Canada, V9E 2E7}\affil[b]{University of British Columbia, 2329 West Mall, Vancouver, Canada, V6T 1Z4}
\affil[c]{University of Victoria, 3800 Finnerty Road, Victoria, Canada, V80 5C2}

\cftpagenumbersoff{figure}
\cftpagenumbersoff{table} 
\begin{document} 
\maketitle

\begin{abstract}

We use telemetry data from the Gemini North ALTAIR adaptive optics system to investigate how well the \textcolor{black}{commands for wavefront correction (both Tip/Tilt and high-order turbulence)} can be forecasted in order to reduce lag error \textcolor{black}{(due to wavefront sensor averaging and computational delays)} and improve delivered image quality. We show that a high level of reduction ($\sim$ 5 for Tip-Tilt and $\sim$ 2 for high-order modes) \textcolor{black}{in RMS wavefront error} can be achieved by using a ``forecasting \textcolor{black}{filter}" based on a linear auto-regressive model with only a few coefficients ($\sim$ 30 for Tip-Tilt and $\sim$ 5 for high-order modes) to complement the existing integral servo-controller. Updating this \textcolor{black}{filter} to adapt to evolving observing conditions is computationally inexpensive and requires less than 10 seconds worth of telemetry data. We also use several machine learning models \textcolor{black}{(Long-Short Term Memory and dilated convolutional models)} to evaluate whether further improvements could be achieved with a more sophisticated non-linear model. Our attempts showed no perceptible improvements over linear auto-regressive predictions, even for large lags where residuals from the linear models are high, suggesting that non-linear wavefront distortions for ALTAIR at the Gemini North telescope may not be forecasted with the current setup.
\end{abstract}


{\noindent \footnotesize\textbf{*}Rehan Hafeez,  \linkable{rhafeez@student.ubc.ca} }

\begin{spacing}{2}   % use double spacing for rest of manuscript

\section{Introduction}
\label{sec:intro}
Atmospheric turbulence causes random distortions in incoming light, putting regions of the incoming wavefronts out of phase. When an object is imaged through atmospheric turbulence, the global tip and tilt of the wavefront make the object appear to move. In contrast, higher-order wavefront distortions significantly degrade the image's resolution. An adaptive optics (AO) system aims to reduce the effect of atmospheric turbulence by calculating these distortions and correcting them in real-time using a deformable mirror (DM), complemented by an additional mirror dedicated to stabilizing the image called a Tip/Tilt mirror (TTM). Image stabilization is especially critical as vibrations originating within the observatory can significantly add to atmospheric turbulence-induced Tip and Tilt.

AO systems have now been deployed in many ground-based astronomical observatories as the improvement in angular resolution \textcolor{black}{and sensitivity afforded by AO} can very significantly improve their scientific output. Many of these systems work in closed-loop, similar to the system depicted in Figure \ref{fig:ao_system}, with a wavefront sensor (WFS) measuring the residual wavefront after correction and a real-time controller (RTC) updating the DM and the TTM in order to keep the residual wavefront as small as possible. WFS measurements are obtained by continuously recording images (frames) on a high-speed camera, and the integration time for each image is typically one millisecond. Key to the performance of the AO system is the time it takes \textcolor{black}{to acquire the WFS measurement} and for the RTC to transfer these images and process them into a set of command signals to control the DM and the TTM. This delay, which is typically also on the order of 1 millisecond, causes the command signals to be slightly stale, resulting in a correction error called lag error. If the atmospheric turbulence could be forecast within this millisecond time scale, then the lag error could be reduced. \textcolor{black}{Reducing the lag error} could greatly improve the imaging performance of AO systems, especially those aimed at faint imaging companions, such as exoplanets in the close vicinity of a bright parent star. \textcolor{black}{For such systems, the lag error is often the main contributor to the loss of contrast due to residual atmospheric turbulence }\cite{Guy18}.

\subsection{Forecasting \textcolor{black}{Filter}}
In most existing systems, the servo-controller is a simple integrator with a gain, which can be adjusted to make the \textcolor{black}{best possible} compromise between increasing the rejection bandwidth (high gain) and reducing the propagation of WFS noise into the command signals (low gain) \cite{Gen94}. In this architecture, no attempt is made at predicting the command signals into the future in order to reduce the lag error \textcolor{black}{(an integrator can be seen as a zero-order predictor)}. Yet, it is reasonable to expect a significant level of correlation between realizations of the atmospheric turbulence separated by only one millisecond. Therefore this correlation can be used to attempt to increase the accuracy of the command signals by forecasting. Several methods have already been proposed to explore this idea and design predictive controllers, either based on statistical models of atmospheric turbulence or purely data-driven. See, e.g. the recent work presented on data-driven predictive control \cite{Haffert21} and references therein.

\begin{figure}
\begin{center}
\begin{tabular}{c}
\includegraphics[width=14.0cm]{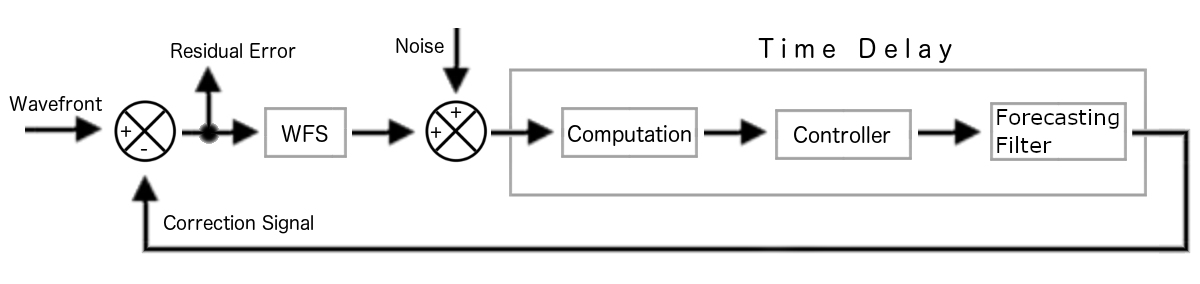}
\end{tabular}
\end{center}
\caption 
{\label{fig:ao_system} Block diagram of a standard AO system, including a forecasting \textcolor{black}{filter} to reduce lag error caused by computational delay.} 
\end{figure}

We propose to add a forecasting \textcolor{black}{filter} to the existing integral control that takes command signals computed from previous frames and forecasts the command for a future frame. We expect such a forecasting \textcolor{black}{filter} would be easy to retrofit as an extension of conventional AO RTCs, where the current forecasting \textcolor{black}{filter} is implicitly the identity operator. Figure \ref{fig:ao_system} shows where in the control system the \textcolor{black}{filter} would be placed.
An ideal forecasting \textcolor{black}{filter} would itself be computationally minimal to keep the additional time delay small. \textcolor{black}{For example, the computational complexity of the forecasting filter could be compared to that of wavefront reconstruction}.

Many previous studies on AO control rely on testing with data obtained by end-to-end AO simulation software implementing a statistical model of the atmospheric turbulence \textcolor{black}{(e.g. \cite{Swanson21}) or by laboratory experiments using artificial turbulence (e.g.  \cite{Haffert21})}. Instead, we design and evaluate our forecasting \textcolor{black}{filter} using a purely data-driven approach based on on-sky AO telemetry data, as outlined in section \ref{sec:data}. \textcolor{black}{AO telemetry data are usually readily available in operating AO systems and have been successfully used in a few previous studies \cite{Poyneer08}\cite{Jensen-Clem19}\cite{vanKooten19}.} In section \ref{sec:timeseriesanalysis}, we examine the command signal as a time series, and in section \ref{sec:models}, we outline the linear forecasting models.

In section \ref{sec:results}, we apply these models to the data and approximate parameters such as how many previous frames are required and what levels of improvement are attainable.  Finally, in section \ref{sec:RA} we explore more complex, non-linear models for forecasting using Neural Networks.

\section{Data}
\label{sec:data}
 
We use AO telemetry data collected on ALTAIR at the Gemini North Telescope between August 2007 and March 2008. Each data set was obtained on a different date, at the beginning of the night, during the daily ``M1 tuning" procedure, where the AO system is locked on a bright natural guide star (NGS) near zenith and persistent wavefront errors corrected by the AO system are offloaded to the primary mirror. \textcolor{black}{The first 20 Zernike modes are continuously offloaded (tip-tilt, defocus, and coma are actually offloaded to M2) at slow speeds ($<$ 1 Hz), so the offload process does not affect the ~millisecond time scale forecasting we are trying to achieve here.} \textcolor{black}{Seeing values for these data are not available; however, it can be expected that they span a range around ~0.5 arcsec, which is the typical median seeing at this site}. In all cases, the AO system was operating at a frame rate of 1kHz, which means that one WFS measurement was obtained and one command vector signal was sent every millisecond. Each telemetry data set, called ``session" hereafter, contains a time series of various real-time data processed by the AO systems, including WFS images, WFS slopes (gradients), reconstructed wavefront errors, and DM/TTM commands. For this work, we are only using the DM/TTM commands.

\begin{figure}
\begin{center}
\begin{tabular}{c}
\includegraphics[width=16.0cm]{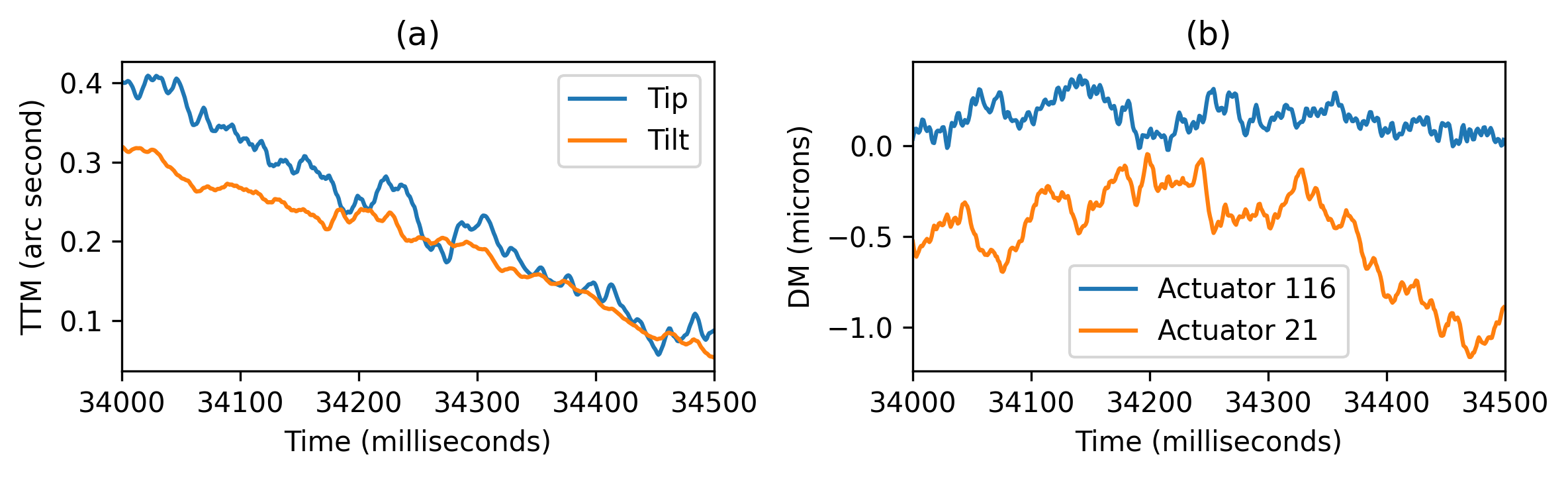}
\end{tabular}
\end{center}
\caption
{\label{fig:raw_timeseries} Command signals as time series data sampled from a M1 tuning. \textbf{a} Tip and Tilt commands for image stabilization.
\textbf{b} Commands to two of the 136 active channels of the DM for high-order correction.}
\end{figure}

In NGS mode, ALTAIR uses a 12x12 Shack-Hartmann WFS to drive a 177-actuator DM plus one Tip-Tilt mirror (TTM) dedicated to image stabilization, at a rate of up to 1 kHz\cite{Sad03}.
Only 136 actuators in the DM are actively controlled while the remaining actuators, located at the edge of the DM, are simply extrapolated.  Commands to the DM actuators are recorded in a 177-element vector in units of microns of wavefront correction. TTM commands are recorded in a 2-element vector in units of arcseconds of image motion on sky. Figure \ref{fig:raw_timeseries} illustrates samples of both types of command signals over 500 milliseconds. \textcolor{black}{For an eight meter telescope, 1 milliarcsecond of image tip or tilt is roughly equivalent to 9.7 nm RMS of wavefront error.}

Session length was limited by buffer capacity, therefore each session contains less than 60 seconds of data. We have chosen sessions with at least 50 seconds of data.
Since these command values are recorded from a real system and analyzed after the fact, we are also limited to evaluating the predictability of these commands. That is, we are not able to test the stability of the system if the proposed forecasting \textcolor{black}{filter} were implemented.

Finally, since for all these sessions the NGS is very bright, the AO loop operates at the maximum possible gain. The correction residuals are largely dominated by lag error and we can therefore assume that the DM and TTM command signals accurately track the incoming atmospheric turbulence with a simple delay. Therefore, for a single frame delay, the optimal command signals at time $t+1$ would be the command signals that were computed by the system for that frame.

\section{Time series Analysis}
\label{sec:timeseriesanalysis}

\begin{figure}
\begin{center}
\begin{tabular}{c}
\includegraphics[width=16.0cm]{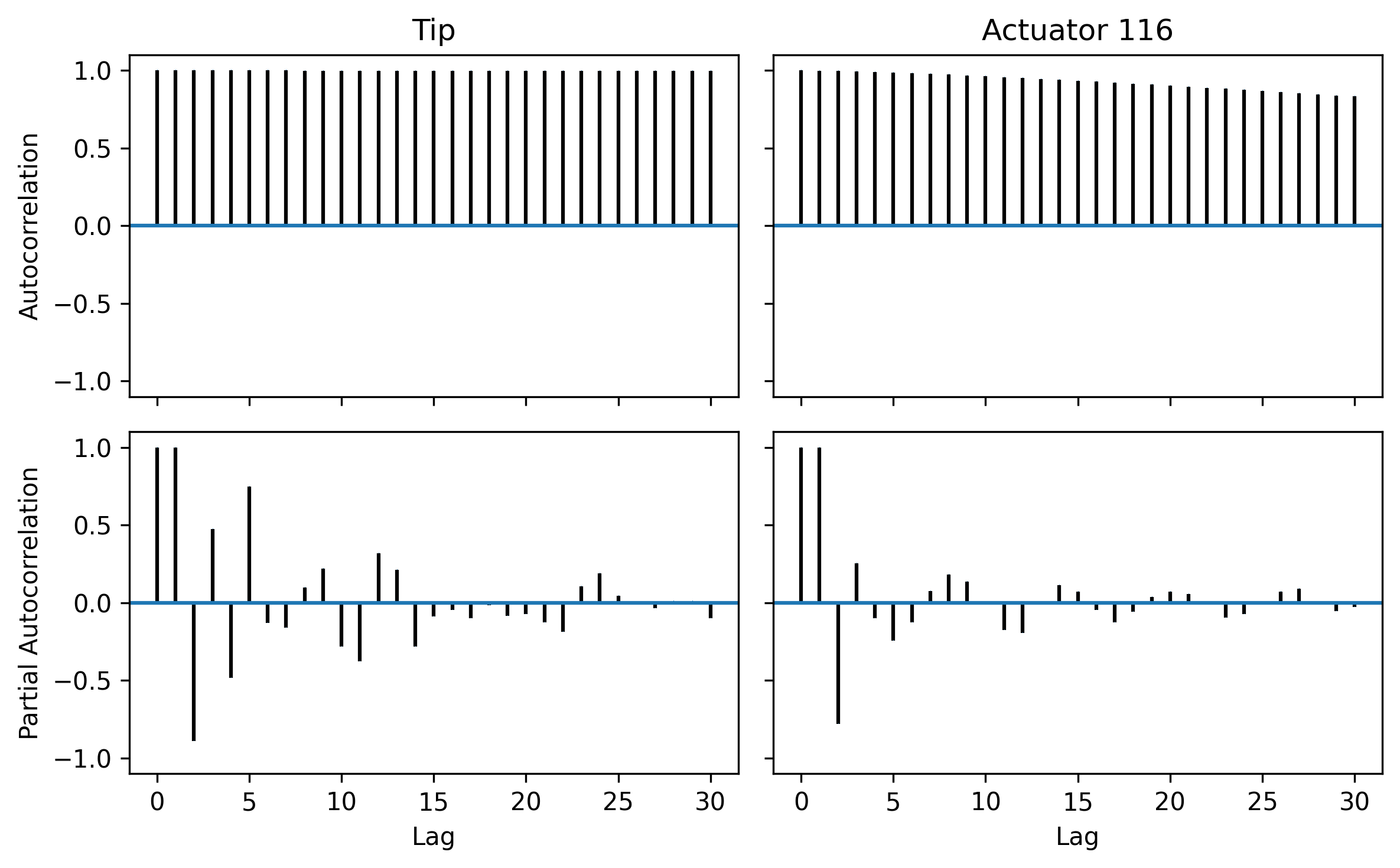}
\end{tabular}
\end{center}
\caption 
{\label{fig:tt_acf} Time series analysis of the Tip channel and Actuator 116. The top row shows Autocorrelations.
Partial autocorrelation shows the impact of each relative frame for forecasting.}
\end{figure} 

We performed exploratory data analysis on a session to understand their general dynamics. The \textcolor{black}{first column in Figure \ref{fig:tt_acf} shows the autocorrelation and partial autocorrelation for the Tip channel for an entire 60 second session. Actuator 116 is shown in the second column and was selected because it is neither in the centre nor towards the edge}.
%This subset of data was not stationary, as determined by the augmented Dickey-Fuller test. To make this subset stationary while maintaining longer-term dynamics, we subtracted the subsequent command at each frame.
%As seen in Figure \ref{fig:tt_acf}a, the oscillations in the autocorrelation function show significant longer-term structure in the subtracted command signal.
As seen in the autocorrelation row in Figure \ref{fig:tt_acf}, there is strong correlation between commands over a long time horizon.
\textcolor{black}{The row of partial autocorrelations suggests that the majority of the signal can be accounted for using only the most recent commands, which is consistent with an autoregressive process. While these partial autocorrelations become less important, they do not completely disappear. This indicates} we can further extend \textcolor{black}{the window of} our model for marginal improvement.

In time series analysis, lag is the term used for the number of previous observations. Since lag could be mistaken for the error caused by computation time in an AO system, we introduce the term Lookback when referring to the number of previous commands the model has access to.

\section{Models}
\label{sec:models}

Our baseline models for both types of data (TTM and DM) are autoregressive models. Forecasts can be expressed by the following equation:

\[\hat{y}_{t+1}^{AR} = \sum_{i=0}^{i=n}w_i y_{t-i} + b \] 

\noindent where the autoregressive forecast ($\hat{y}_{t+1}^{AR}$) is dependent on the relative time index (\textit{t}), the Lookback number (\textit{n}), a constant bias term (\textit{b}), and fixed parameters of the model (\textit{w}). 

As a metric for evaluating our models, we will use the Root Mean Square Error (RMSE) between the next frame forecast \textit{\{$\hat{y}_{t+1}$\}} and the actual observed ground truth \textit{\{${y}_{t+1}$\}}. 
Our baseline is the current ALTAIR system, which we call Echo, where the absence of any forecasting returns the full lag error; that is $\hat{y}_{t+1}^{Echo} = y_{t}$.

\subsection{Image Stabilization}
\label{sec:imagestab}
Image stabilization is controlled using Tip and Tilt \textcolor{black}{Mirror (TTM)}. Dynamic image motion is caused by atmospheric turbulence but also by other effects within the telescope, such as wind shake and vibrations caused by electro-mechanical systems. 

Our aim is to make forecasts for both the Tip and Tilt channels \textcolor{black}{separately}. We define two \textcolor{black}{types for forecasting}; 
\textcolor{black}{models of the first type use a single channel and assumes Tip and Tilt are independent, the other allows for linear interactions between both channels.}
%one pair with single-channel inputs, and the other pair using both channels as inputs to forecasting a single channel output.
In the \textcolor{black}{independent type forecasting, a single channel is used as input to forecast the command corresponding to that channel.
Models using the dependent type of forecasting have access to both channels, and therefore double the number of parameters needed to define them compared to the independent type.} %single channel models, the two channels are treated as independent, whereas in the double channel models, we do not make this assumption.
For example, a Single$_{tip}$ model trained with a Lookback of 50 will have 50 parameter coefficients\textcolor{black}{;} a Double$_{tip}$ model with the same Lookback \textcolor{black}{will have 100 parameter coefficients}: 50 derived from each of the Tip and Tilt channels and makes a forecast for the Tip channel. \textcolor{black}{The model parameters are found with Ordinary Least Squares using a small amount of data from the session.}

\subsection{High-Order Turbulence Mitigation}

The actuators of the DM are arranged in a square lattice with an annular shape. The uncontrolled actuators are located around the edge and one in the center of the array (\textcolor{black}{caused by }telescope central obscuration). For simplicity, we use separate linear models for each of the 136 controllable actuators. \textcolor{black}{ These 136 separate forecasts are compiled into a single forecast for the DM.}

We also examine the quality of forecasts using modes of the large-scale patterns of deformation in the mirror\textcolor{black}{, such as Zernike modes}. Low order modes, such as Defocus and Astigmatisms, contain most of the variance of the atmospheric turbulence \textcolor{black}{and have a higher signal to noise ratio than higher order modes. While we see in section \ref{sec:dmmode} that the amount of Lookback required for a given improvement may decrease with mode index, the method we present here keeps the amount of Lookback consistent for each mode}.

We use the ALTAIR system's modal basis, which spans the entire wavefront sub-space spanned by the 136 controlled DM actuators. This orthonormal modal basis is constructed such that each mode fits as closely as possible the corresponding Zernike polynomial. Converting into modal space is done via a 136x136 conversion matrix, which is available in an ALTAIR configuration file. Again, one model was fitted to each mode, meaning the forecast requires combining many small models together before being converted back into actuator commands. 
The benefit of forecasting commands using modes is that a few modes can be used to forecast the majority of the variation while ignoring the noisier high order modes. \textcolor{black}{This is more important where computational limits exist and implementing one forecasting filter per DM actuator is not feasible.} Also, while the ALTAIR modes are not rigorously statistically independent, the low order modes are very close to the Karhunen-Loeve modes for the Kolmogorov turbulence, making the underlying assumption of statistical independence reasonable.

\section{Results}
\label{sec:results}
Our results are separated into two parts, TTM forecasting and DM forecasting. Figures in this section use data collected on 11th January 2008 (11Jan08), unless otherwise noted. We will show that this is a representative session because, although specific wavefront corrections may vary across different dates, the general trends in improvement remain consistent. 

\subsection{Tip and Tilt Forecasting}
\label{sec:tt_results}

\begin{figure}[ht]
\begin{center}
\begin{tabular}{c}
\includegraphics[width=16.0cm]{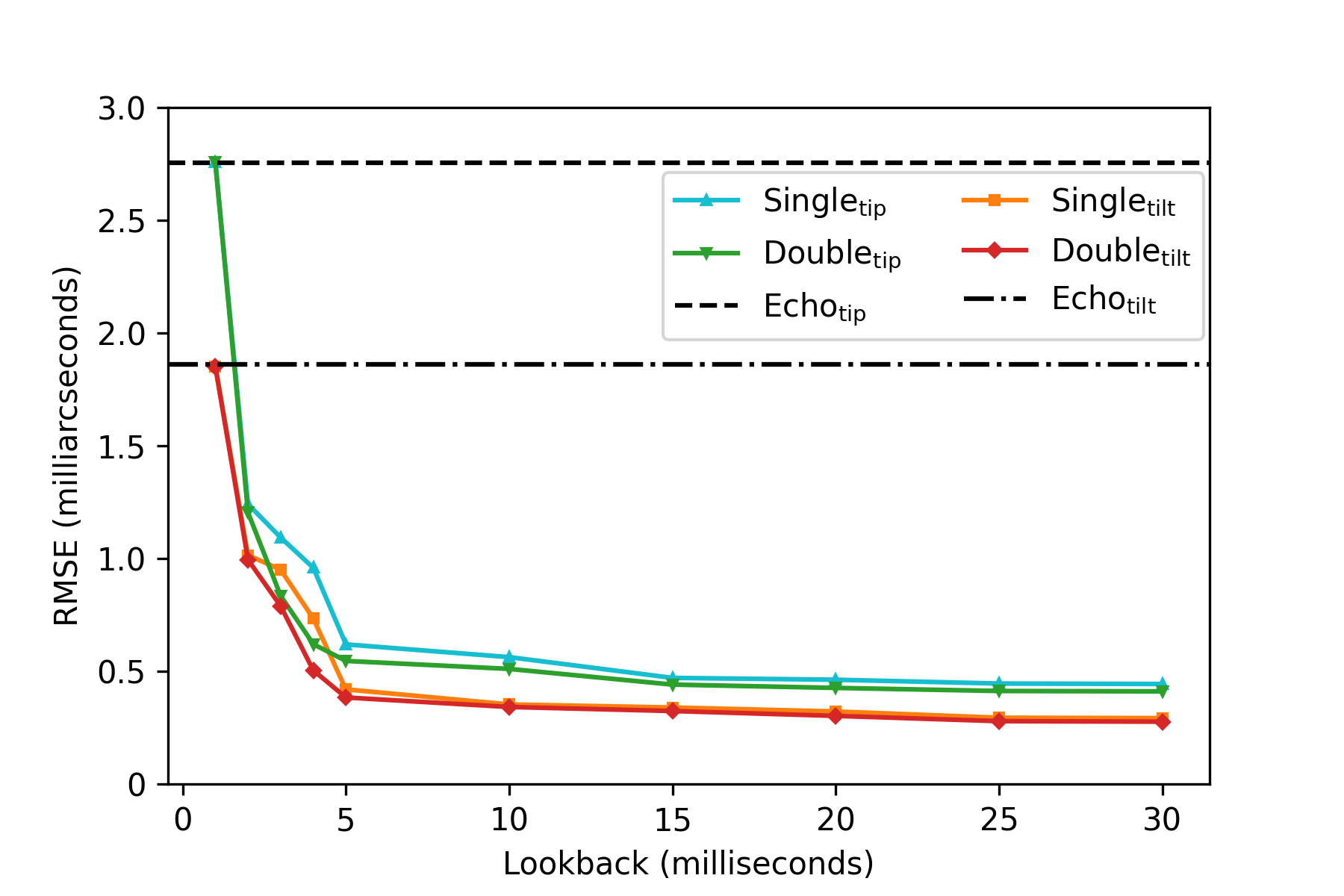}
\end{tabular}
\end{center}
\caption 
{ \label{fig:ttlag}
Forecasting error as a function of Lookback.}
\end{figure} 

Lookback is the number of previous frames we give a model when fitting. This AO system uses a 1kHz frame rate, so a Lookback of \textcolor{black}{20} corresponds to \textcolor{black}{20} milliseconds. Each model was evaluated with a 90/10 train/test split. Figure \ref{fig:ttlag} shows that increasing the Lookback reduces the modeling error. Our baseline of using no forecasting \textcolor{black}{filter} (Echo) is shown as a dashed horizontal line for each channel. These values are stationary because the Echo forecast is equivalent to having a Lookback of 1. We see that most of the improvement is achieved for a Lookback as low as 5, and using a Lookback greater than \textcolor{black}{30} has marginal improvements. \textcolor{black}{Moreover,} by using both channels for forecasting, a slight reduction in error can be achieved.  

\begin{figure}[ht]
\begin{center}
\begin{tabular}{c}
\includegraphics[width=16cm]{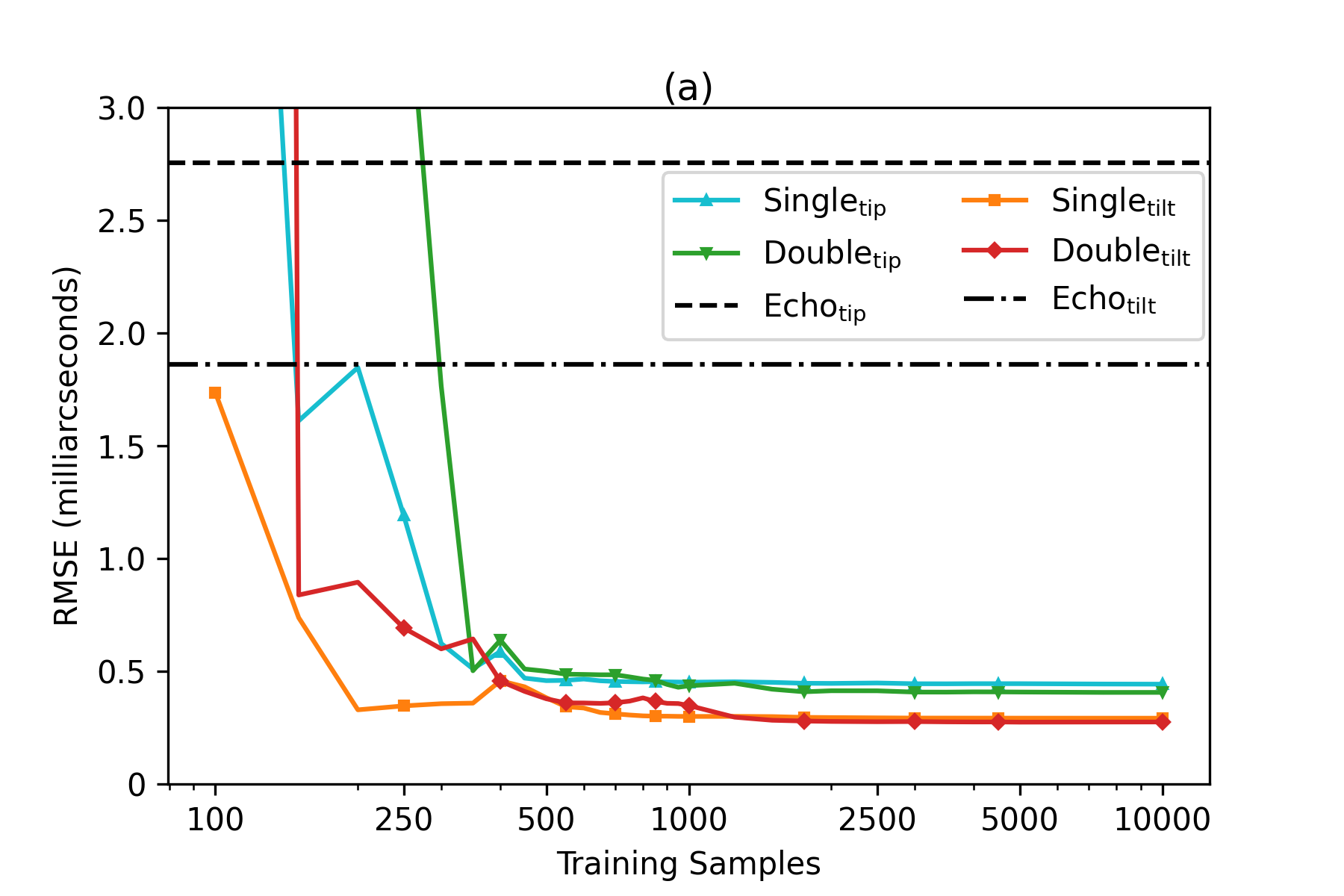}
\end{tabular}
\end{center}
\caption 
{ \label{fig:ttsamples}
Forecasting error as function of the number of training samples}
\end{figure}

Given a Lookback of \textcolor{black}{30}, we then test how many samples are required to fit the models. To obtain these results, the test set was held constant as the final 10\% of the session, and the samples used for fitting the models immediately precede the test set. Therefore, using fewer samples does not introduce a delay between the two sets of data, so models can be directly compared. Figure \ref{fig:ttsamples} shows
\textcolor{black}{the RMS tip-tilt error (RMSE) as a function of training data. Each point shows a model fit with the amount of training samples and evaluated on the test set. This figure shows}
that 10,000 samples \textcolor{black}{are} more than sufficient, representing 10 seconds worth of data. Double$_{tip}$ models require more samples for training, as they have more degrees of freedom. While we use 10,000 samples for training going forward, similar results can be achieved using fewer samples. \textcolor{black}{Echo$_{tip}$ and Echo$_{tilt}$ provide a reference RMSE of these channels for this part of the session.}

\begin{figure}[ht]
\begin{center}
\begin{tabular}{c}
\includegraphics[width=16cm]{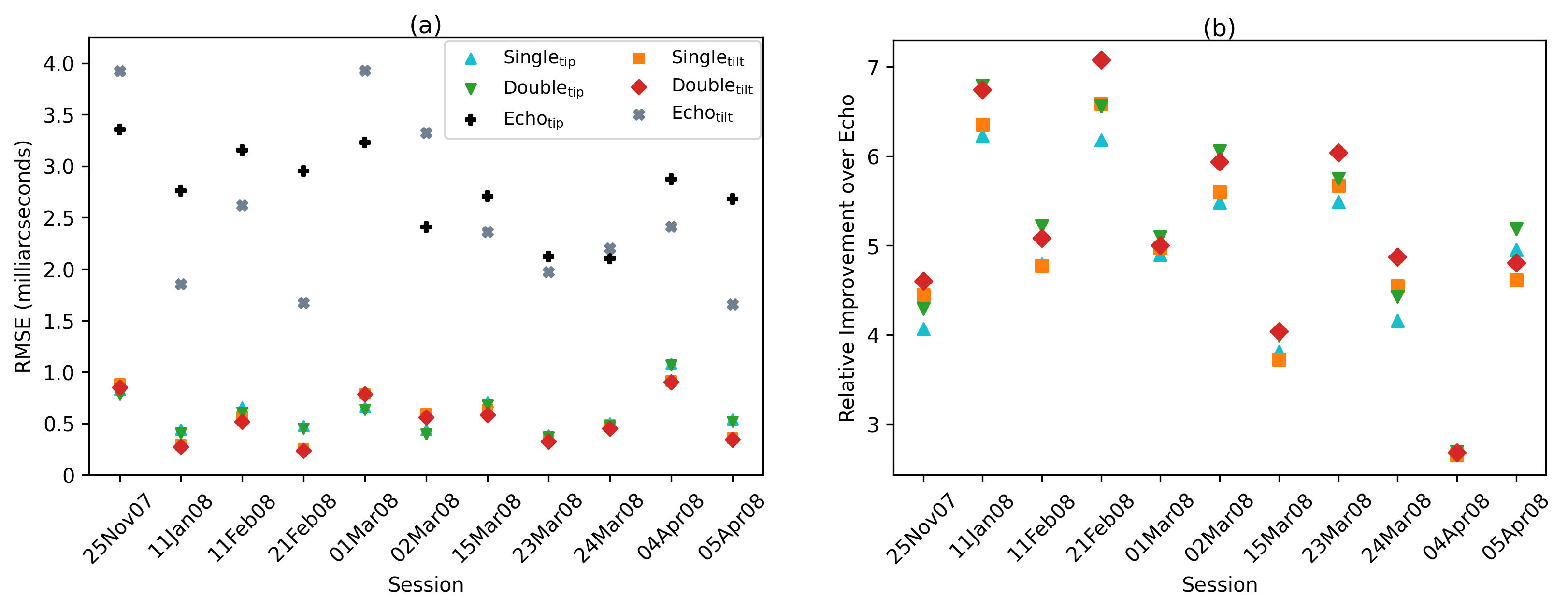}
\end{tabular}
\end{center}
\caption 
{ \label{fig:ttsessions}
Forecasting error for different sessions. \textbf{a} Full vertical range; \textbf{b} \textcolor{black}{Relative improvement over echo} }
\end{figure} 

Figure \ref{fig:ttsessions} shows the improvement of our models compared to Echo for multiple sessions. \textcolor{black}{We fit models for each channel for each session and evaluated them on a testing set composed of the final 10\% of the session. Models were fit with} a Lookback of \textcolor{black}{30}, with 10,000 samples. Figure \ref{fig:ttsessions}a  shows significant improvements over Echo. Figure \ref{fig:ttsessions}b \textcolor{black}{shows a factor improvement of our method above Echo and more clearly} illustrates that the Double$_{tip}$ models consistently perform better than the Single$_{tip}$ models.

\begin{figure}[H]
\begin{center}
\begin{tabular}{c}
\includegraphics[width=16.0cm]{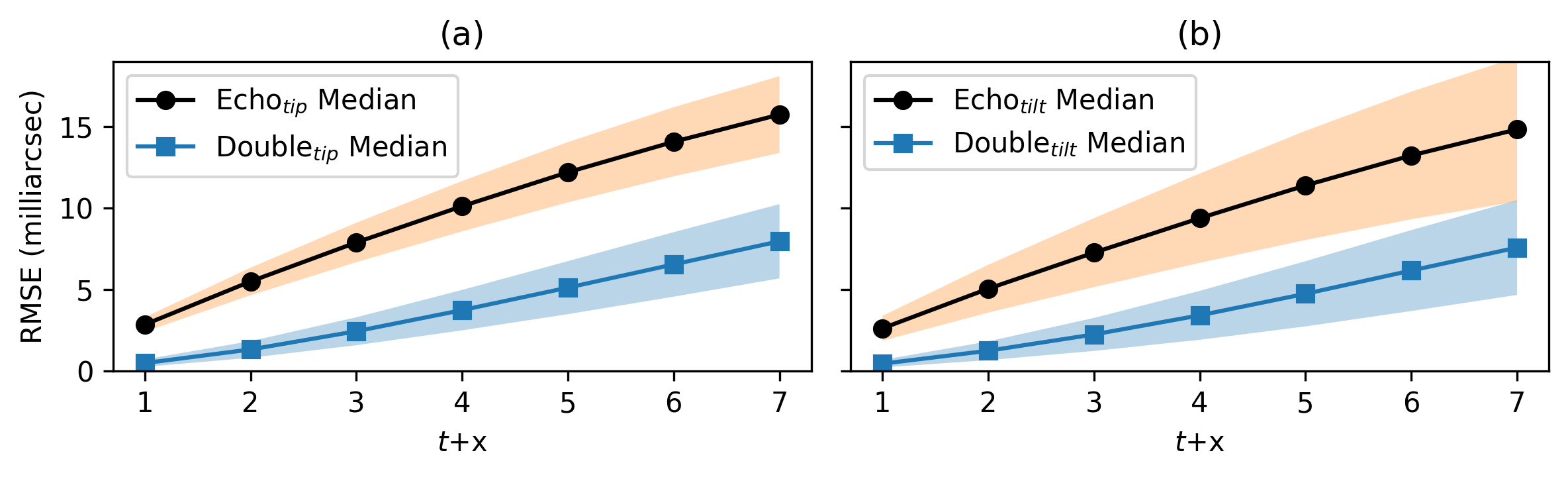}
\end{tabular}
\end{center}
\caption 
{ \label{fig:ttplusx}
TTM forecast error as a function of AO time delay, expressed in number of frames for Tip \textbf{(a)} and Tilt \textbf{(b)} channels. The median values are calculated from 32 sessions and the filled area is one standard deviation.}
\end{figure}

Forecasting for delays larger than one frame increases the forecasting error. However, in Figure \ref{fig:ttplusx} we see our methodology can still improve the lag error compared to the Echo baseline. This trend is observed over multiple sessions, with increased variation observed at larger time delays.
The ability to forecast for longer time delays is particularly interesting, as it opens the possibility to "slow down" the AO system in order to increase the exposure time on the WFS, allowing the observer to use fainter sources, and therefore increasing the portion of the sky available for NGS AO observations \textcolor{black}{\cite{Doelman20}}. 

\subsection{High-order turbulence forecasting}
For the high order turbulence corrected by the DM, we test our linear forecasting model in actuator space (section \ref{sec:dmactuator}), and in modal space (section \ref{sec:dmmode}). These models are the same linear autoregressive models outlines in section \ref{sec:models}.

\begin{figure}
\begin{center}
\begin{tabular}{c}
\includegraphics[width=16.0cm]{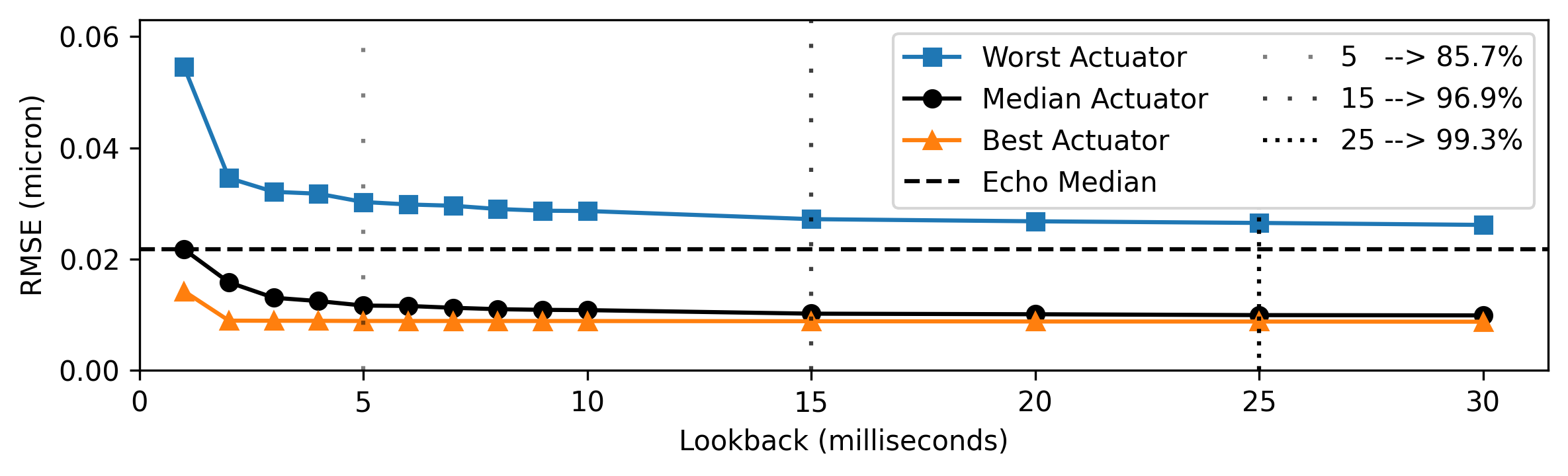}
\end{tabular}
\end{center}
\caption 
{ \label{fig:dmactuators}
Forecasting error as a function of Lookback for high order turbulence using actuator commands as inputs. Each session has 136 actuators, shown here are the two extremes and the median forecasting quality. Vertical dashed lines show the relative improvement of the median compared to the best observed improvement (i.e., at Lookback = 30).}
\end{figure}

\begin{figure}[b]
\begin{center}
\begin{tabular}{c}
\includegraphics[width=16.0cm]{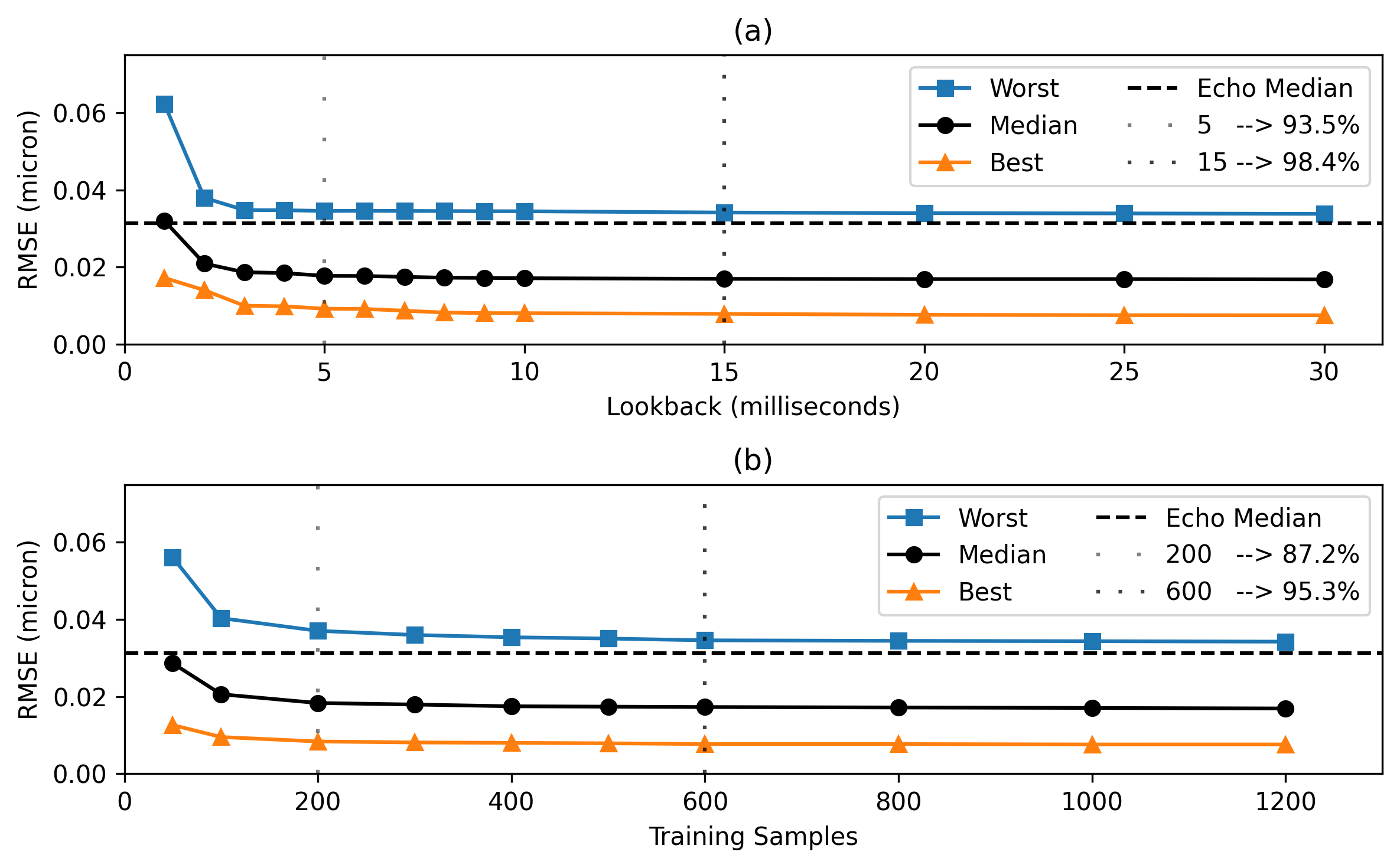}
\end{tabular}
\end{center}
\caption 
{ \label{fig:dmmulti}
Forecasting error of multiple sessions as a function of (\textbf{a}) Lookback and (\textbf{b}) number of training samples for high order turbulence using actuator commands as inputs. Each session is summarized by its median performance, and we have plotted the best, worst, and median sessions. Vertical dashed lines show improvement relative to the best observed median improvement. \textbf{a} used 10,000 samples for training, and \textbf{b} used a Lookback of 20.}
\end{figure}

\subsubsection{Actuator Space Forecasting}
\label{sec:dmactuator}
In actuator space we use raw actuator commands to forecast subsequent actuator commands. We test models with a Lookback of up to 30 milliseconds on a single session.
Figure \ref{fig:dmactuators} shows that the forecast improvement plateaus beyond a Lookback of 20, and that over $80\%$ of the improvement can be captured with a Lookback of 5 \textcolor{black}{for this session}. The distribution of actuator errors is a function of where they are located in the mirror, with central and edge actuators accounting for the highest improvements. This makes sense because those actuators are controlled by fewer and noisier WFS sub-apertures, and therefore carry more measurement noise. Because of this variation, we show the best and worst performing actuator forecasts as well as the median. The best and worst actuators are selected by their forecasting quality at Lookback = 1. Vertical dashed lines show the improvement of the median relative to its lowest observed RMSE, which occurs at Lookback = 30.

To illustrate that 11Jan08 is a representative session of the range of turbulence conditions, we show in Figure \ref{fig:dmmulti} the trends for multiple sessions. Figure \ref{fig:dmmulti} shows median session forecast quality and we have plotted the best, worst, and median session qualities. Although there is left-over variability within the RMSE values for each session, the trends across sessions are comparable. To help with visual clarity, \textcolor{black}{we have only shown the median Echo}.

\begin{figure}[b]
\begin{center}
\begin{tabular}{c}
\includegraphics[width=16.0cm]{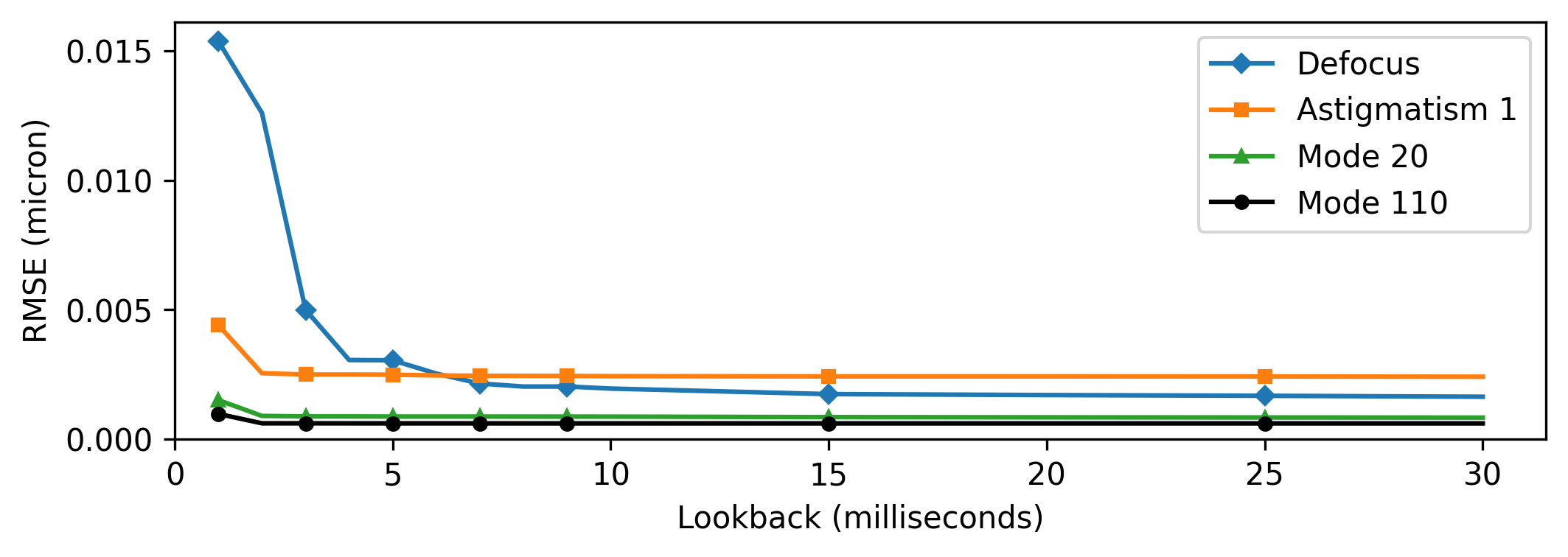}
\end{tabular}
\end{center}
\caption 
{ \label{fig:dmmodes}
Forecasting error as a function of Lookback in modal space. Four modes are selected for plotting. Defocus is Mode 2, Astigmatism 1 is Mode 3. 10,000 samples were used for training.}
\end{figure}

\subsubsection{Modal Space Forecasting}
\label{sec:dmmode}
In modal space, actuator commands are converted into modal commands with a conversion matrix. The models in this section then take modal inputs and forecast the modal representation of commands for the subsequent frame.
Unlike actuator data, amplitudes vary between modes. As shown in Figure \ref{fig:dmmodes}, Defocus (the first high-order mode) largely dominates, with amplitudes of subsequent modes decreasing as mode number increases, as expected in a Kolmogorov turbulence. For all modes except Defocus, we find that the forecasting error plateaus within a Lookback of 5. Due to its much higher amplitude, Defocus benefits from a longer Lookback.

\begin{figure}
\begin{center}
\begin{tabular}{c}
\includegraphics[width=16.0cm]{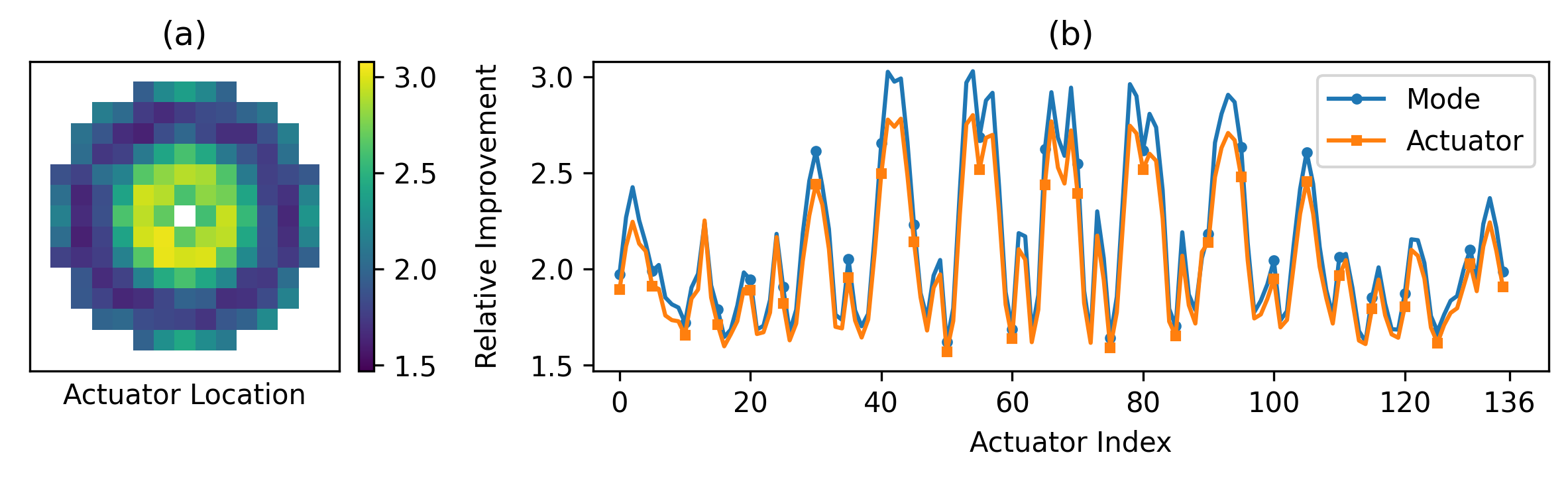}
\end{tabular}
\end{center}
\caption 
{ \label{fig:dmresults}
Comparison of DM forecast by method. \textbf{a} Relative improvement of Modal forecasts mapped to actuator locations over the Echo forecasts. The central pixel is occluded by the telescope.
\textbf{b} The relative improvement over an Echo forecast at each actuator by each method.}
\end{figure} 

Forecasting using either modes or actuators provides an improvement over Echo. We compare each method by looking at the relative improvement over the Echo baseline. First, we convert the modal forecast into actuator space. For each actuator we derive a single value score representing the relative improvement by dividing the RMSE of the Echo forecast by the RMSE of the respective method.
Figure \ref{fig:dmresults}a shows where on the actuator map these improvements occur. These values are from the modal forecast, but the actuator forecast has a nearly identical distribution. The center pixel is blank as it is obscured by the telescope. In Figure  \ref{fig:dmresults}b the relative improvement by each method is unravelled into actuator index.
Forecasting in modal space results in a lower forecasting error compared to forecasting actuators directly. Since forecasting using modes can influence the entire array, we find that $\sim$ 50\% of our observed improvement is from the first 5 modes \textcolor{black}{starting with Defocus. Therefore, the computational requirement could be reduced if needed by forecasting only a limited number of modes. It is worth noting that here we have restricted to one filter per actuator (or per mode). Grouping neighbouring actuators using a spatial filter as explained in \cite{Jensen-Clem19} might allow taking advantage of local spatial correlations in the wavefront and thus improve the actuator space forecasting.}

\section{Residual Analysis}
\label{sec:RA}
While our proposed autoregressive models provide a significant improvement over Echo, we still observe an increasing error when forecasting further into the future (Figure \ref{fig:ttplusx}). This observation is consistent for both the Tip/Tilt commands and high-order modes. We expect that some non-linear interactions might be present in the time series, which the autoregressive models lack capacity to capture.   
These non-linear interactions would be present in the residual error of our linear forecasts. In this section, we explore whether we could take advantage of these non-linear interactions to improve our forecasts using Neural Network (NN) models.

Using NN architectures to forecast the atmospheric turbulence in real time has already been explored \cite{Swanson21,Wong21}, however, as far as we know, none of these approaches have been tested on sky. Also, these studies do not explicitly separate linear and non-linear forecasting components, making it difficult to evaluate the real benefit of implementing a more complex modelling approach. 

\textcolor{black}{As not all NN architectures can take advantage of the time ordering in the data, we focus our attention on  three NN architectures designed to incorporate states while processing the next inputs in the sequence. We settled on three specific recurrent and convolutional architectures, reviewed in} a recent survey by Lim et al. of machine learning algorithms for time series forecasting\cite{Lim21}. 
We limit our analyses to the Tip channel for TTM, and to Mode 22 for DM. 
\textcolor{black}{We focus our analysis on residuals from forecasts at $t+1$ to provide a direct comparison to our linear model, and at $t+7$ in order to reasonably maximize the error from non-linear effects.}
%Since the linear residuals at $t+1$ are very small, we focus our analysis on residuals from forecasts at $t+7$ in order to reasonably maximize the error from non-linear effects.

\subsection{Methodology}
We define the residual error between linear forecasts and true actuator commands at \textcolor{black}{$t+{\tau}$ as a new time series, where $\tau$ represents the forecasted time.}
The NNs take actuator commands from previous frames and fits them to the residual of the corresponding linear forecast at \textcolor{black}{$t+\tau$}. This method would extract any potential non-linear trends from the actuator commands that the linear model was unable to capture.
The time series residual is extracted from the 11Jan08 session. For the Tip channel analysis, we use forecasts made by the Double$_{tip}$ model, which is fit with a Lookback of \textcolor{black}{30}. Similarly, for Mode 22, we use forecasts made with a linear model fit with a Lookback of \textcolor{black}{30}. These forecasts are subtracted from the true actuator commands, resulting in a time series with approximately 60,000 frames. The NNs are trained using the first 90\% of the residual time series.

\begin{figure}
\begin{center}
\begin{tabular}{c}
\includegraphics[width=10.0cm]{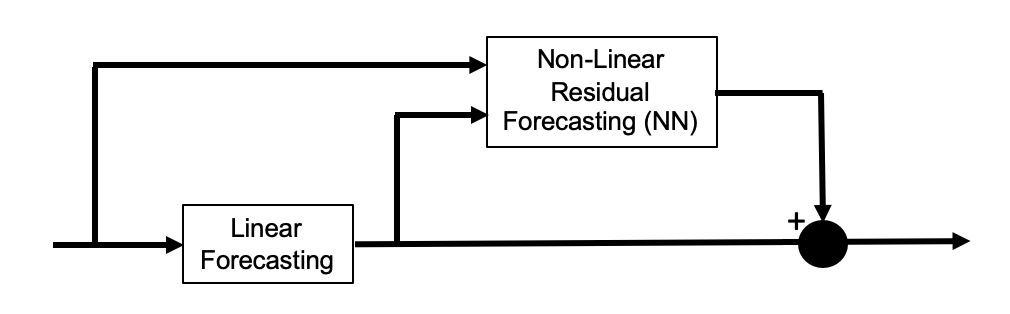}
\end{tabular}
\end{center}
\caption 
{\label{fig:NNPredictor} Block diagram of a forecasting \textcolor{black}{filter} with linear and residual forecasting \textcolor{black}{filter}.} 
\end{figure}

We generate a corrected actuator forecast by adding the residual forecasted by the NNs to the corresponding linear forecast. 
If the corrected forecast yields a significantly lower RMSE than the corresponding autoregressive forecast alone, then we can estimate the degree of non-linearity within the time series. Such information could inform the development of more sophisticated models that could be implemented in real time. For this analysis, the new ``forecasting \textcolor{black}{filter}" in Figure \ref{fig:ao_system} is extended to include a ``non-linear residual forecasting \textcolor{black}{filter}" as shown in Figure \ref{fig:NNPredictor}

\subsection{Neural Network Architectures}
We limit our evaluation of NN architectures to three types, following Lim et. al\cite{Lim21}, namely (1) Recurrent Neural Networks, implemented as a Long Short-Term Memory (LSTM) network, (2) Attention Mechanisms, implemented as an LSTM with an Attention layer and (3) Dilated Convolutional Networks, implemented as the WaveNet architecture.

Both (1) and (2) rely on LSTM units, which are regularly used to model stateful sequences due to their ability to handle long-range dependencies \textcolor{black}{\cite{Hochreiter97} by incorporating a hidden state and a gated memory cell from the previous inputs. Whereas traditional sequence-to-sequence architectures based on recurrent neural networks need the whole sequence, ultimately keeping only the last compressed state, attention mechanism \cite{Bahdanau15} add the extra capacity to look (attend) at all previous hidden states of the the sequence in the Lookback window, assigning weights to the most relevant ones for the prediction.}

%Attention mechanisms cluster together temporal features by dynamically assigning weights to input features. This allows the network to 'pay attention' to more significant time steps in the Lookback window\cite{Lim21}. 

The WaveNet architecture was first developed as a generative model for raw audio \cite{WaveNet}, \textcolor{black}{and is based on convolutional neural network, handling the whole input sequence, but also masking the future inputs to preserve temporal structure from the input data}. This popular network architecture has been shown to perform well not only in audio generation, but also language processing and time series forecasts \cite{Lim21}.

Each architecture went through a process of hyperparameter optimization by training each model with a range of values, and in each case we used a \textcolor{black}{90/10} validation split. For (1) and (2) we found a stable loss plateau after 500 epochs using a batch size of 2048, with the Adam optimizer set with a learning rate of 1e-5 \textcolor{black}{for the Tip channel analysis, and 1e-6 for Mode 22}. In the case of (3), the same loss plateau is achieved using a batch size of 1024, with the \textcolor{black}{Adam optimizer set to a learning rate of 1e-5} after 25 epochs.  

\subsection{Results}
In Table 1, we present the results of the NN correction for Tip and Mode 22 \textcolor{black}{for both $t+1$ and $t+7$}, which we derive by adding the forecasted residual to the corresponding linear model forecast. From this we calculate the RMSE of the corrected forecasts against the true actuator values. Due to the stochastic nature of the machine learning algorithms used, we present the mean value of the RMSE alongside the standard deviations derived from 10 different seed values on the same session.
For comparison, we also present the RMSE of the baseline linear model without any non-linear correction.  \textcolor{black}{Since the Linear Baseline is deterministic and we are using the same session to isolate the NN initialization variation, the Linear Baseline does not have error bars.}

\begin{table}[H]
\begin{center}
\begin{tabular}{||c| c c |c c||} 
 \hline
 \multirow[c]{3}{*}{Model} & \multicolumn{2}{c|}{\textit{t+1}} & \multicolumn{2}{c|}{\textit{t+7}}\\ &\textit{tip} RMSE & \textit{Mode 22} RMSE & \textit{tip} RMSE & \textit{Mode 22} RMSE\\   &(mas) & (nm) & (mas) & (nm) \\[0.5ex] 
 \hline\hline
 Simple LSTM & 0.414 ± 0.002 & 0.830 ± 0.0003 & 8.681 ± 0.006 & 6.147 ± 0.007\\ 
 Attention LSTM & 0.417 ± 0.006 & 0.831  ± 0.0004 & $\mathbf{8.672}$ ± $\mathbf{0.007}$ & $\mathbf{6.143}$ ± $\mathbf{0.002}$\\
 Wavenet & 0.516 ± 0.137 & 0.830 ± 0.0001 & 8.699 ± 0.026 & 6.150 ± 0.001\\ 
 Linear Baseline & $\mathbf{0.411}$ & $\mathbf{0.830}$ & 8.680 & 6.157\\ [1ex] 
 \hline
\end{tabular}
\end{center}
\caption{Summary of NN architectures and performance compared to linear baseline.}
\label{table:1}
\end{table}

\textcolor{black}{In most cases the} corrected forecasts made by the NN models approach but do not exceed the performance of the autoregressive models. \textcolor{black}{When the corrected forecasts exceed the performance of the autoregressive model, the improvements are very marginal (percent of a mas or percent of a nm) and come at great computational cost.} We find no evidence that additional information can be \textcolor{black}{practically} extracted from these models to aid in forecasting.  

\section{Conclusion and Discussion}
\label{sec:discussion}
In this paper, we have used on-sky AO telemetry data acquired on ALTAIR at Gemini North to show that a simple data-driven autoregressive forecaster is very efficient at forecasting AO correction to mitigate latency of one or more frames. This seems to be especially true for Tip and Tilt, possibly because these signals have sinusoidal vibration components that are easy to model. Over several different nights, we have also found that the Tip/Tilt RMSE could be improved by a factor of $\sim$ 5 by forecasting one frame ahead, using a Lookback of 30 frames. However, increasing Lookback time continues to marginally improve forecasting accuracy, since lower temporal frequencies can be better modelled, at the cost of increasing the complexity of the forecaster (i.e., number of operations to execute in real time). We have also found that jointly forecasting Tip and Tilt results in slightly better results than considering the two channels independently. 

For high-order modes (Defocus and up), we have found that one frame ahead linear forecasting can reduce the RMSE by a factor of $\sim$ 2. We have found that forecasting in modal space resulted in a slightly lower error. However, the real advantage of modal space forecasting is that only forecasting the first few ($\sim$ 5) modes is enough to achieve most of the improvement.

\textcolor{black}{We note that the wavefront error improvements we are finding are better than the theoretical maximum improvement (~1.3) that has been derived in reference \cite{Doelman20}. This is likely because reference \cite{Doelman20} only consider atmospheric effects, whereas our data have significant non-atmospheric components, certainly on T/T, likely on defocus and possibly other modes, mainly due to vibrations.  We are finding a higher improvement because stationary vibrations are easier to forecast.}

Our results also suggest the possibility of forecasting ahead by more than one frame, opening the possibility to decrease the AO frame rate and therefore increase sky coverage. \textcolor{black}{Moreover, with a Lookback of 30 frames, 30 multiplies and add are required for each channel, which means that the computational complexity for the forecasting filter is much less than that of the wavefront computation process (which requires a number of multiplies and adds equal to the number of WFS slopes), which suggests that the forecasting filter could be easily retrofitted in the existing real-time controller.}

Finally, we have found that using Neural Networks to complement linear forecasting did not seem to bring any additional benefit, despite trying three different architectures. This suggests that the most efficient way to model the atmospheric turbulence to be corrected in our setup is with an autoregressive model, and that residuals from such a model are random noise that cannot be forecast in any way. \textcolor{black}{These residuals could include non-linear effects in the atmospheric turbulence or in the Altair system itself.}

These conclusions are only valid for the ALTAIR AO system and only for the data sets we had access to, with the evaluated Neural Network models. They would need to be verified on other systems and at other sites. However, we have highlighted the potential benefit of a simple data-driven linear forecasting model. Based on the ALTAIR data we analyzed, we would propose using a 30 coefficient filter for Tip and Tilt, and a 5 coefficient filter for each of the DM actuator channels (modal coefficients are not available in real time in ALTAIR). This would result in a modest increase in computational load (i.e., compared to the wavefront reconstruction process), and therefore could potentially be implemented easily in the existing Real Time Controller. Further more, updating the coefficients to reflect current observing conditions can be done by a soft real time process looking at \textcolor{black}{past} telemetry data \textcolor{black}{just like loop gains are currently optimized in Altair}.
% \disclosures 

\section* {Code, Data, and Materials Availability} 
Code and data may be made available upon request.

%%%%% References %%%%%

\bibliography{report}   % bibliography data in report.bib
\bibliographystyle{spiejour}   % makes bibtex use spiejour.bst

%%%%% Biographies of authors %%%%%

\vspace{2ex}\noindent\textbf{Rehan Hafeez}  is an engineering physics student at the University of British Columbia whose experience includes testing and prototyping of electromechanical systems, neural network optimization, and most recently data science and machine learning with the National Research Council of Canada. Driven by a desire to optimize how we interface with our environment, Rehan’s goal is to develop sustainable technologies to support and advance research.

\vspace{1ex}
\noindent Biographies and photographs of the other authors are not available.

\end{spacing}
\end{document}